\documentclass[conference]{IEEEtran}
\usepackage{amsfonts}
\IEEEoverridecommandlockouts

\ifCLASSINFOpdf
\else
\fi
\usepackage{epsfig}
\usepackage{graphicx}
\usepackage{psfig}
\usepackage{epsf}
\usepackage[cmex10]{amsmath}
\usepackage{subfigure}
\usepackage{booktabs}
\usepackage{fancyhdr}
\usepackage{slashbox}
\usepackage{caption2}   

\begin{document}
\title{Resource Allocation for Secure MISO-NOMA Cognitive Radios Relying on SWIPT}
\author{Fuhui Zhou$^{ * \ddag} $, Zheng Chu$^\dagger$, Haijian Sun$^* $,  Victor C. M. Leung$^\S$\\
 $^*$Utah State University, Utah, USA,  $^{\ddag}$Nanchang University, Nanchang, China,\\ $^\dagger$University of Surrey, London, U.K,
$^\S$The University of British Columbia, Vancouver, Canada\\

Email: \emph{\{zhoufuhui@ieee.org,  andrew.chuzheng7@gmail.com, h.j.sun@ieee.org,  vleung@ece.ubc.ca\}}
\thanks{The research was supported by the National Science Foundation under the grant EARS-1547312, the National Natural Science Foundation of China (61701214, 61661028, 61631015, and  61561034), the Young Natural Science Foundation of Jiangxi Province (20171BAB212002), the China Postdoctoral Science Foundation (2017M610400) and The Postdoctoral Science Foundation of Jiangxi Province(2017KY04, 2017RC17).}}
\maketitle
\begin{abstract}
Cognitive radio (CR) and non-orthogonal multiple access (NOMA) are two promising technologies  in the next generation wireless communication systems. The  security of a NOMA CR network (CRN) is important but lacks of study. In this paper, a multiple-input single-output NOMA CRN relying on simultaneous wireless information and power transfer is studied. In order to improve the security of both the primary and secondary network, an artificial noise-aided cooperative jamming scheme is proposed. Different from the most existing works, a power minimization problem is formulated under a practical non-linear energy harvesting model. A suboptimal scheme is proposed to solve this problem based on semidefinite relaxation and successive convex approximation. Simulation results show that the  proposed cooperative jamming scheme is efficient to achieve secure communication and NOMA outperforms the conventional orthogonal multiple access in terms of the power consumption.
\end{abstract}
\begin{IEEEkeywords}
Cognitive radio, non-orthogonal multiple access, physical-layer security, simultaneous wireless information and power transfer.
\end{IEEEkeywords}
\IEEEpeerreviewmaketitle
\section{Introduction}
\IEEEPARstart{T}{HE} next generation wireless communication systems call for advanced communication techniques that can achieve high spectral efficiency (SE)  and provide massive connectivity in support of the escalating high data rate requirements imposed by the unprecedented proliferation of mobile devices \cite{J. G. Andrews}. Cognitive radio (CR) and non-orthogonal multiple access (NOMA) are promising due to their high SE and the capability of providing massive connectivity. CR enables the secondary network to access the spectrum band of the primary network as long as the interference caused to the primary network is tolerable \cite{S. Haykin}. Different from orthogonal multiple access (OMA), NOMA has the potential advantages in SE and user connectivity by using non-orthogonal resources at the cost of the receiver's implementation complexity \cite{Z. Ding}, \cite{R. Q. Hu1}. It is envisioned that the application of NOMA into CR networks (CRNs) can significantly improve SE and user connectivity \cite{Y. Liu}, \cite{Z. Zhang}.

Meanwhile, the next generation wireless communication systems also need energy-efficient techniques due to the ever increasing greenhouse gas emission concerns and explosive proliferation of power-limited devices, e.g., sensors and mobile phones.  To that end, simultaneous wireless information and power transfer (SWIPT) has drawn great attentions \cite{X. Lu}. It can simultaneously transmit information and achieve energy harvesting (EH). Particularly, radio frequency (RF) signals carry not only  information, but also are identified as energy sources for EH. Compared with the conventional EH techniques, such as wind charging, SWIPT can provide a stable and controllable
power for energy-limited devices. Thus, in NOMA CRNs with power-limited devices, it is of significant importance to study the application of SWIPT into NOMA CRNs.

However, due to the broadcasting nature of NOMA as well as CR and the dual function of RF signals \cite{F. Zhou2}-\cite{Y. Zhang1}, NOMA CRNs relying on SWIPT are vulnerable to eavesdropping. Malicious energy harvesting receivers (EHRs) may exist and intercept the confidential information transmitted to the primary users (PUs) and the secondary users (SUs) \cite{F. Zhou2}. Thus, it is vital to improve the security of NOMA CRNs relying on SWIPT. As an alternative to the traditional cryptographic techniques, physical-layer security exploits the physical characteristics (e.g., multipath fading, propagation delay, etc.) of wireless channels to achieve secure communications. It was shown that the secrecy rate of wireless communication systems is limited by the channel state information (CSI) \cite{Y. Zou1}. In order to improve the secrecy rate, multiple antennas, cooperative relay, jamming and artificial noise (AN)-aided techniques have been applied \cite{X. Chen}, \cite{Z. Chu}.

Many investigations have been conducted to improve the security of the conventional OMA systems and initial efforts have been made to study secure transmission in NOMA systems \cite{Y. Zhang1}, \cite{Y. Li}-\cite{B. He}. However, to authors' best knowledge, few investigations have been conduced for improving the security of NOMA CRNs relying on SWIPT. The existing works for OMA systems relying on SWIPT can be categorized into two research lines based on the energy harvesting model, namely, the linear EH model \cite{F. Zhou2}, \cite{D. W. K. Ng1}, \cite{C. Xu} and the non-linear EH model \cite{E. Boshkovska3}, \cite{E. Boshkovska}-\cite{K. Xiong}. In \cite{F. Zhou2}, the authors studied robust beamforming design problems in MISO CRNs relying on SWIPT based on the linear EH model. Under this model, the harvested power linearly increases with the input power. The authors in \cite{D. W. K. Ng1} established a multiple-objective optimization framework in MISO CRNs relying on SWIPT. It was shown that there exist multiple tradeoffs, such as the tradeoff between the harvesting energy and the secrecy rate. In \cite{C. Xu}, the secure transmission problems were extended into multiple-input multiple-output (MIMO) CRNs. Obviously, the linear EH model is ideal due to the practical non-linear end-to-end power conversation circuit \cite{E. Boshkovska3}, \cite{E. Boshkovska}-\cite{K. Xiong}. Recently, the authors in  \cite{E. Boshkovska3}, \cite{E. Boshkovska}-\cite{K. Xiong}  proposed a non-linear EH model and studied resource allocation problems. In \cite{E. Boshkovska3} and \cite{E. Boshkovska}, beamforming design problems were studied in MISO systems relying on SWIPT based on the proposed non-linear EH model. It was shown that the harvesting energy achieved under the non-linear EH model may be higher than that obtained under the linear EH model. These problems were extended into MIMO systems relying on SWIPT in \cite{E. Boshkovska2} and \cite{K. Xiong}.

However, the beamforming schemes proposed in \cite{F. Zhou2}, \cite{E. Boshkovska3}, \cite{D. W. K. Ng1}-\cite{K. Xiong} are inappropriate to NOMA CRNs relying on SWIPT since NOMA is very different from OMA. Although the works in \cite{Y. Zhang1}, \cite{Y. Li}-\cite{B. He} studied resource allocation problems in NOMA systems, these resource allocation schemes are  unadaptable to NOMA CRNs relying on SWIPT. The reasons  are from the following two perspectives. On the one hand, they were proposed for the conventional NOMA systems that do not need to consider the interference between the primary network and the secondary network. On the other hand, SWIPT was not applied and the EH requirement was not considered.

In this paper, in order to improve the security of the primary network, an AN-aided cooperative scheme is proposed. By using this scheme, the cognitive base station (CBS) transmits a jamming signal to cooperate with the primary base station (PBS) for improving the security of the PUs. As a reward, the secondary network is granted to access the frequency bands of the primary network and provide SWIPT services both for the SUs and for the EHRs in the secondary network. The transmission beamforming and AN covariance matrix are jointly optimized to minimize the total transmission power of the network while the secrecy rate and the EH requirement are guaranteed. Simulation results show that  our proposed cooperative scheme is efficient and NOMA outperforms OMA in terms of the power consumption.

The rest of this paper is organized as follows. Section II presents the system model. The AN-aided beamforming design problem is formulated in Section III. Section IV presents simulation results. The paper concludes with Section V.

\emph{Notations:} Boldface capital letters and boldface lower case letters represent matrices and vectors, respectively. The Hermitian (conjugate) transpose, trace, and rank of a matrix \textbf{A} are represented respectively by $\mathbf{A^H}$, Tr$\left(\mathbf{A}\right)$ and Rank$\left(\mathbf{A}\right)$. $\mathbf{I}$ denotes the identity matrix. The conjugate transpose of a vector $\mathbf{x}$ is denoted by $\mathbf{x}^\dag$. $\mathbf{C}^{M\times N}$ denotes a $M$-by-$N$ dimensional complex matrix set. $\mathbf{A}\succeq \mathbf{0} \left(\mathbf{A}\succ \mathbf{0}\right)$ represents that $\mathbf{A}$ is a Hermitian positive semidefinite (definite) matrix. $\mathbb{H}^N$ and $\mathbb{H}_+^{N}$ denote a $N$-by-$N$ dimensional Hermitian matrix set and a Hermitian positive semidefinite matrix set, respectively. ${\left\|  \cdot  \right\|}$ denotes the Euclidean norm of a vector. The absolute value of a complex scalar is denoted by ${\left| \cdot \right|}$. $\mathbf{x} \sim {\cal C}{\cal N}\left( {\mathbf{u},\mathbf{\Sigma } }\right)$ means that $\mathbf{x}$ is a random vector and follows a complex Gaussian distribution with mean $\mathbf{u}$ and covariance matrix $\mathbf{\Sigma }$. $\mathbb{E}[ \cdot ]$ denotes the expectation operator.
\section{System Model}
\begin{figure}[!t]
\centering
\includegraphics[width=3.1 in]{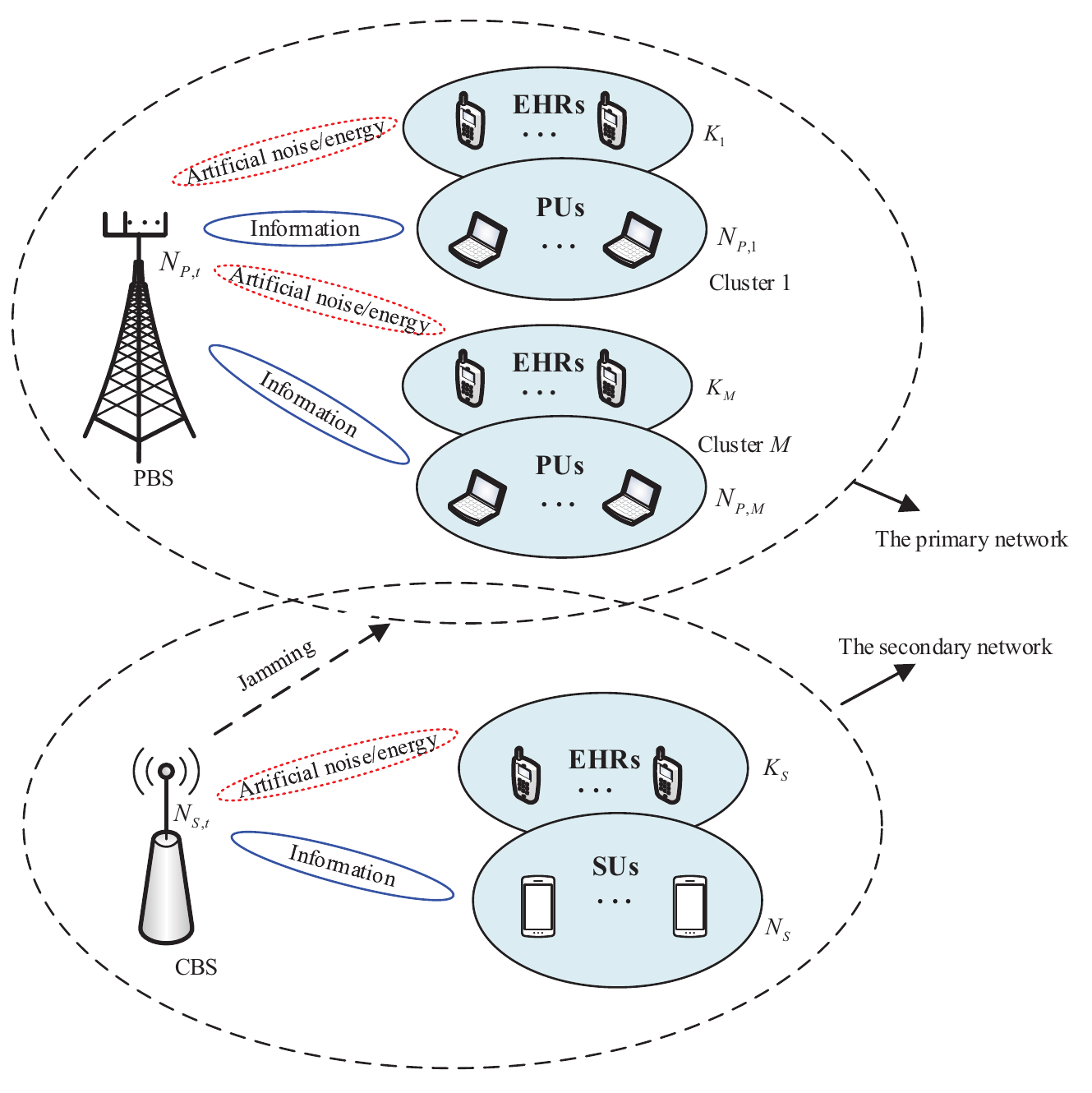}
\caption{The system model.} \label{fig.1}
\end{figure}
A downlink MISO NOMA CR network with SWIPT is considered in Fig. 1. In the primary network, multicast communications are exploited since they can provide high SE and massive connectivity but PUs' receivers are simple, which cannot perform SIC. This scenario is widely encountered, for example in Internet of Things, wireless sensor networks, and cellular network \cite{X. Chen}. In the secondary network, NOMA is applied since it can achieve high power transfer efficiency and SUs can perform successive interference cancellation (SIC) \cite{Y. Liu}. In this case, the PBS broadcasts information to PUs in $M$ clusters and simultaneously transfers energy to EHRs. In the secondary network, the CBS provides SWIPT service to $K_s$ EHRs and $N_s$ SUs by using NOMA. The PBS is equipped  with $N_{P,t}$ antennas and the CBS is equipped with $N_{S,t}$ antennas. All the PUs, SUs and EHRs are equipped with a single antenna. Due to the broadcasting characteristic of NOMA and the dual function of RF signals, the EHR may eavesdrop and intercept the information transmitted by the PBS and the CBS. It is assumed that EHRs in each network can only intercept confidential information from the same network and PUs in each cluster are respectively wiretapped by EHRs in the same cluster. For example, PUs in the $m$th cluster, where $m\in {\cal M}$ and $ {\cal M}\buildrel \Delta \over = \left\{ {1,2, \cdots ,M} \right\}$, are wiretapped by the $k$th EHR in the $m$th cluster, where $k\in {\cal K}_m$ and $ {\cal K}_m\buildrel \Delta \over = \left\{ {1,2, \cdots ,K_m} \right\}$. $K_m$ is the number of EHRs and $N_{P,m}$ is the number of PUs in the $m$th cluster. In order to improve the security of the primary network, an An-aided cooperative scheme is applied. Using this scheme, the CBS of Fig. 1 transmits a jamming signal to the primary network for improving the security of the PUs. As a reward, the primary network allows the secondary network to operate on its frequency bands. All the channels involved are assumed to be flat fading channels. It is assumed that all CSI is assumed to be perfect \cite{Y. Zhang1}, \cite{Y. Li}, \cite{M. Tian}. The performance achieved under this assumption can be used as a bound analysis and provides meaningful insights into the design of MISO NOMA CRNs.

Let ${y_{P,m,i}}$, ${y_{S,j}}$, ${y_{E,m,k}}$ and ${y_{E,l}}$ denote the signal received  at the $i$th PU in the $m$th cluster and the $j$th SU, and the EH signal at the $k$th EHR in the $m$th cluster and the $l$th EHR in the secondary network, respectively, where $i\in {\cal N}_{P,m}$, ${\cal N}_{P,m}=\left\{ {1,2, \cdots ,N_{P,m}} \right\}$; $j\in {\cal N}_{s}$, ${\cal N}_{s}=\left\{ {1,2, \cdots ,N_{s}} \right\}$ and $l\in {\cal K}_{s}$, ${\cal K}_{s}=\left\{ {1,2, \cdots ,K_s} \right\}$. These signals are respectively given as
\begin{subequations}
\begin{align}\label{27}\ \notag
&{y_{P,m,i}} = \mathbf{h}_{P,m,i}^\dag \left[ {\sum\limits_{m = 1}^M \left({{\mathbf{w}_{p,m}}{s_{p,m}}}  + {\mathbf{v}_{p,m}}\right)} \right]  \\
&\ \ \ \ \ \ \ \ \ \ + \mathbf{f}_{S,m,i}^\dag \left( {\sum\limits_{j = 1}^{N_s} {{\mathbf{w}_{s,j}}{s_{s,j}}}  + {\mathbf{v}_s}} \right) + {n_{P,m,i}}, \\ \notag
&{y_{S,j}} = \mathbf{q}_{P,j}^\dag \left[ {\sum\limits_{m = 1}^M \left({{\mathbf{w}_{p,m}}{s_{p,m}}}  + {\mathbf{v}_{p,m}}\right)} \right] \\
&\ \ \ \ \ \ \ \ + \mathbf{h}_{S,j}^\dag \left( {\sum\limits_{j = 1}^{N_s} {{\mathbf{w}_{s,j}}{s_{s,j}}}  + {\mathbf{v}_s}} \right) + {n_{S,j}}, \\ \notag
&{y_{E,m,k}} = \mathbf{g}_{E,m,k}^\dag \left[ {\sum\limits_{m = 1}^M \left({{\mathbf{w}_{p,m}}{s_{p,m}}}  + {\mathbf{v}_{p,m}}\right)} \right]\\
&\ \ \ \ \ \ \ \ \ \  + \mathbf{f}_{E,m,k}^\dag \left( {\sum\limits_{j = 1}^{N_s} {{\mathbf{w}_{s,j}}{s_{s,j}}}  + {\mathbf{v}_s}} \right), \\ \notag
&{y_{E,l}} = \mathbf{q}_{E,l}^\dag \left[ {\sum\limits_{m = 1}^M \left({{\mathbf{w}_{p,m}}{s_{p,m}}}  + {\mathbf{v}_{p,m}}\right)} \right] \\
&\ \ \ \ \ \ \ \ \ + \mathbf{g}_{E,l}^\dag \left( {\sum\limits_{j = 1}^{N_s} {{\mathbf{w}_{s,j}}{s_{s,j}}}  + {\mathbf{v}_s}} \right),
\end{align}
\end{subequations}
where $\mathbf{h}_{P,m,i}\in {\mathbf{C}^{{N_{P,t}} \times 1}}$ and $\mathbf{f}_{S,m,i}\in {\mathbf{C}^{{N_{S,t}} \times 1}}$ are the channel vector between the PBS and the $i$th PU and that between the CBS and the $i$th PU in the $m$th cluster, respectively; $\mathbf{q}_{P,j}\in {\mathbf{C}^{{N_{P,t}} \times 1}}$ and $\mathbf{h}_{S,j}\in {\mathbf{C}^{{N_{S,t}} \times 1}}$ denote the channel vector between the PBS and the $j$th SU and that between the CBS and the $j$th SU, respectively; $\mathbf{g}_{E,m,k}\in {\mathbf{C}^{{N_{P,t}} \times 1}}$ and $\mathbf{f}_{E,m,k}\in {\mathbf{C}^{{N_{S,t}} \times 1}}$ are the channel vector between the PBS and the $k$th EHR and that between the CBS and the $j$th EHR in the $m$th cluster, respectively; $\mathbf{q}_{E,l}\in {\mathbf{C}^{{N_{P,t}} \times 1}}$ and $\mathbf{g}_{E,l}\in {\mathbf{C}^{{N_{S,t}} \times 1}}$ represent the channel vector between the PBS and the $l$th EHR and that between the CBS and the $l$th EHR in the secondary network, respectively. Still regarding to $\left(1\rm{a}\right)$, ${s_{p,m}} \in {\mathbf{C}^{{1} \times 1}}$ and $\mathbf{w}_{p,m}\in {\mathbf{C}^{N_{P,t} \times 1}}$ are the confidential information-bearing signal for PUs in the $m$th cluster and the corresponding beamforming vector, respectively; ${s_{s,j}} \in {\mathbf{C}^{{1} \times 1}}$ and $\mathbf{w}_{s,j}\in {\mathbf{C}^{N_{S,t} \times 1}}$ represent the confidential information-bearing signal for the $j$th SU and the corresponding beamforming vector, respectively; ${\mathbf{v}_{p,m}}$ and ${\mathbf{v}_s}$ denote the noise vector artificially generated by the PBS and the CBS for improving the security of these two networks. Without loss of generality, it is assumed that $\mathbb{E}[ {{{\left| {s_{p,m}} \right|}^2}} ] = 1$ and $\mathbb{E}[ {{{\left| {s_{s,j}} \right|}^2}} ] = 1$. It is also assumed that $\mathbf{v}_{p,m} \sim {\cal C}{\cal N}\left( {0,\mathbf{\Sigma}_{p,m} } \right)$ and $\mathbf{v}_s \sim {\cal C}{\cal N}\left( {0,\mathbf{\Sigma}_s}  \right)$, where $\mathbf{\Sigma}_{p,m}$ and $\mathbf{\Sigma}_s$ are the AN covariance matrix to be designed. In $\left(1\right)$, ${n_{P,m,i}} \sim {\cal C}{\cal N}\left( {0,\sigma _{P,m,i}^2} \right)$ and ${n_{S,j}} \sim {\cal C}{\cal N}\left( {0,\sigma _{S,j}^2} \right)$ respectively denote the complex Gaussian noise at the $i$th PU in the $m$th cluster and the $l$th SU.

Let $\mathbf{W}_{p,m}=\mathbf{w}_{p,m}\mathbf{w}_{p,m}^\dag$; $\mathbf{W}_{s,j}=\mathbf{w}_{s,j}\mathbf{w}_{s,j}^\dag$; $\mathbf{H}_{P,m,i}=\mathbf{h}_{P,m,i}\mathbf{h}_{P,m,i}^\dag$; $\mathbf{F}_{S,m,i}=\mathbf{f}_{S,m,i}\mathbf{f}_{S,m,i}^\dag$; $\mathbf{Q}_{P,j}=\mathbf{q}_{P,j}\mathbf{q}_{P,j}^\dag$; $\mathbf{H}_{S,j}=\mathbf{h}_{S,j}\mathbf{h}_{S,j}^\dag$; $\mathbf{G}_{E,m,k}=\mathbf{g}_{E,m,k}\mathbf{g}_{E,m,k}^\dag$; $\mathbf{F}_{E,m,k}=\mathbf{f}_{E,m,k}\mathbf{f}_{E,m,k}^\dag$; $\mathbf{Q}_{E,l}=\mathbf{q}_{E,l}\mathbf{q}_{E,l}^\dag$ and $\mathbf{G}_{E,l}=\mathbf{g}_{E,l}\mathbf{g}_{E,l}^\dag$. Based on $\left(1\right)$, the secrecy rate of the $i$th PU in the $m$th cluster and the secrecy rate of the $j$th SU, denoted by $R_{P,m,i}$ and ${R_{S,j}}$, respectively, can be expressed as
\begin{subequations}
\begin{align}\label{27}\ \notag
&{R_{P,m,i}} = \log \left( {\frac{{{\Gamma _{P,m,i}}}}{{{\Gamma _{P,m,i}} - \text{Tr}\left( {{\mathbf{W}_{p,m}}{\mathbf{H}_{P,m,i}}} \right)}}} \right) \\
&\ \ \ \ - \mathop {\max }\limits_{k \in {{\cal K}_m}} \log \left( {\frac{{{\Gamma _{E,m,k}} + \sigma _{E,m,k}^2}}{{{\Gamma _{E,m,k}} - \text{Tr}\left( {{\mathbf{W}_{p,m}}{\mathbf{G}_{E,m,k}}} \right) + \sigma _{E,m,k}^2}}} \right).\\ \notag
&{R_{S,j}}\\
& \left\{ \begin{array}{l}
{=\log _2}\left( {\frac{{{\Gamma _{S,j}}}}{{{\Gamma _{S,j}} - \text{Tr}\left( {{\mathbf{W}_{s,j}}{\mathbf{H}_{S,j}}} \right)}}} \right)\\
 \ \ \ \ - \mathop {\max }\limits_{l \in \cal L} {\log _2}\left( {\frac{{{\Lambda _{E,l,j}}}}{{{\Lambda _{E,l,j}} -\text{Tr}\left( {{\mathbf{W}_{s,j}}{\mathbf{G}_{E,l}}} \right)}}} \right),j = {N_s}, \\
=\mathop {\min }\limits_{z \in \left\{ {j,j + 1,{N_s}} \right\}} {\log _2}\left( {\frac{{{\Lambda _{S,j,z}}}}{{{\Lambda _{S,j,z}} - \text{Tr}\left( {{\mathbf{W}_{s,j}}{\mathbf{H}_{S,z}}} \right)}}} \right)\\
 - \mathop {\max }\limits_{l \in \cal L} {\log _2}\left( {\frac{{{\Lambda _{S,l,j}}}}{{{\Lambda _{S,l,j}} - \text{Tr}\left( {{\mathbf{W}_{s,j}}{\mathbf{G}_{E,l}}} \right)}}} \right),j = 1, \cdots ,{N_s-1}
\end{array} \right.
\end{align}
\end{subequations}
where $\Gamma _{P,m,i}$, $\Gamma _{E,m,k}$, $\Gamma _{S,j}$, $\Lambda _{E,l,j}$, $\Lambda _{S,j,z}$ and $\Lambda _{S,l,j}$ are given as $\left(3\right)$ at the top of the next page. Without loss of generality, it is assumed that $\left\| {{\mathbf{h}_1}} \right\| \leq \left\| {{\mathbf{h}_2}} \right\|\leq\cdots \leq \left\| {{\mathbf{h}_{N_s}}} \right\|$. Similar to \cite{Y. Zhang1}, \cite{Y. Li}-\cite{B. He}, it is assumed that the EHR in the secondary network has decoded  SU's $j$'s message before it decodes the SU's $i$'s  message, $j<i$. This
overestimates the interception capability of EHRs and results in the worst-case secrecy rate of SUs. This conservative assumption was  used  in \cite{Y. Zhang1}, \cite{Y. Li}-\cite{B. He}.
\begin{figure*}[!t]
\normalsize
\begin{subequations}
\begin{align}\label{27}\
&{\Gamma _{P,m,i}} = \text{Tr}\left\{ {\left[ {\sum\limits_{m = 1}^M {\left( {{\mathbf{W}_{p,m}} + {\mathbf{\Sigma} _{p,m}}} \right)} } \right]{\mathbf{H}_{P,m,i}} + \left( {\sum\limits_{j = 1}^{{N_s}} {{\mathbf{W}_{s,j}}}  + {\mathbf{\Sigma} _s}} \right){\mathbf{F}_{S,m,i}}} \right\} + \sigma _{P,m,i}^2,\\
&{\Gamma _{E,m,k}} = \text{Tr}\left\{ {\left[ {\sum\limits_{m = 1}^M {\left( {{\mathbf{W}_{p,m}} + {\mathbf{\Sigma} _{p,m}}} \right)} } \right]{\mathbf{G}_{E,m,k}} + \left( {\sum\limits_{j = 1}^{N_s} {{\mathbf{W}_{s,j}}}  + {\mathbf{\Sigma} _s}} \right){\mathbf{F}_{E,m,k}}} \right\},\\
&{\Gamma _{S,j}} = \text{Tr}\left\{ {\left[ {\sum\limits_{m = 1}^M {\left( {{\mathbf{W}_{p,m}} + {\mathbf{\Sigma} _{p,m}}} \right)} } \right]{\mathbf{Q}_{P,j}} + \left( {{\mathbf{W}_{s,j}} + {\mathbf{\Sigma} _s}} \right){\mathbf{H}_{S,j}}} \right\} + \sigma _{S,j}^2,\\
&{\Lambda _{E,l,j}} = \text{Tr}\left[ {\left[ {\sum\limits_{m = 1}^M {\left( {{\mathbf{W}_{p,m}} + {\mathbf{\Sigma} _{p,m}}} \right)} } \right]{\mathbf{Q}_{E,l}} + \left( {{\mathbf{W}_{s,j}} + {\mathbf{\Sigma} _s}} \right){\mathbf{G}_{E,l}}} \right] + \sigma _{E,l}^2,\\
&{\Lambda _{S,j,z}} = \text{Tr}\left\{ {\left[ {\sum\limits_{m = 1}^M {\left( {{\mathbf{W}_{p,m}} + {\mathbf{\Sigma} _{p,m}}} \right)} } \right]{\mathbf{H}_{P,z}} + \left( {\sum\limits_{u = j}^{{N_s}} {{\mathbf{W}_{s,u}}}  + {\mathbf{\Sigma} _s}} \right){\mathbf{H}_{S,z}}} \right\} + \sigma _{S,z}^2,\\
&{\Lambda _{s,l,j}} = \text{Tr}\left\{ {\left[ {\sum\limits_{m = 1}^M {\left( {{\mathbf{W}_{p,m}} + {\mathbf{\Sigma} _{p,m}}} \right)} } \right]{\mathbf{Q}_{E,l}} + \left( {\sum\limits_{\upsilon  = j}^{{N_s}} {{\mathbf{W}_{s,\upsilon }}}  + {\mathbf{\Sigma} _s}} \right){\mathbf{G}_{E,l}}} \right\} + \sigma _{E,l}^2.
\end{align}
\end{subequations}
\hrulefill \vspace*{4pt}
\end{figure*}

In this paper, a practical non-linear EH model is adopted. According to \cite{E. Boshkovska}-\cite{K. Xiong}, the harvesting power of EHRs, denoted by ${\Phi _{E,{\rm A}}}$ can be formulated as:
\begin{subequations}
\begin{align}\label{27}\
&{\Phi _{E,{\rm A}}} = \left( {\frac{{{\psi _{E,{\rm A}}} - P_{E,{\rm A}}^{\max }{\Psi _{E,{\rm A}}}}}{{1 - {\Psi _{E,{\rm A}}}}}} \right),\\
&{\psi _{E,{\rm A}}} = \frac{{P_{E,{\rm A}}^{\max }}}{{1 + {e^{ - {a_{E,{\rm A}}}\left( {{\Gamma _{E,{\rm A}}} - {b_{E,{\rm A}}}} \right)}}}},\\
&{\Psi _{E,{\rm A}}} = \frac{1}{{1 + {e^{{a_{E,{\rm A}}}{b_{E,{\rm A}}}}}}},
\end{align}
\end{subequations}
where $A$ is the set of EHRs in the primary network and the secondary network, namely, $A = {A_1} \cup {A_2}$, and ${A_1} = \mathop  \cup \limits_{m \in \cal M} {\cal K}_m$, $m\in{\cal M}$, ${A_2} =  {\cal K}_{s}$; $a_{E,{\rm A}}$ and $b_{E,{\rm A}}$ represent parameters that reflect the circuit specifications, such as the resistance, the capacitance and diode turn-on voltage \cite{E. Boshkovska}; $P_{E,{\rm A}}^{\max }$ is the maximum harvested power of EHRs when the EH circuit is saturated. In $\left(4\rm{b}\right)$, ${\Gamma _{E,{\rm A}}} $ is the received RF power at EHRs; ${\Gamma _{E,{\rm A}}}= {\Gamma _{E,m,k}}$ when EHRs are in the primary network and ${\Gamma _{E,{\rm A}}}= \Lambda _{s,l,1}-\sigma _{E,l}^2$ when EHRs are in the secondary network. Note that the noise power is ignored since it is small compared to the RF signal power \cite{D. W. K. Ng1}-\cite{K. Xiong}.

\section{AN-aided Beamforming Design}
\subsection{Problem Formulation}
In order to minimize the total transmit power,  the beamforming and the AN covariance of the PBS and the CBS are jointly optimized under constraints of  the secrecy rate of PUs and SUs, the interference power caused to PUs and the EH requirement of EHRs. The power minimization problem is formulated as $\text{P}_{{1}}$ in the following.
\begin{subequations}
\begin{align}\label{27}\
 & \text{P}_{{1}}: \mathop {\min }\limits_{\scriptstyle{\mathbf{W}_{p,m}},{\mathbf{\Sigma} _{p,m}}\hfill\atop
\scriptstyle{\mathbf{W}_{s,j}},{\mathbf{\Sigma} _s}\hfill} \ {\text{Tr}\left[ {\sum\limits_{m = 1}^M {\left( {{\mathbf{W}_{p,m}} + {\mathbf{\Sigma} _{p,m}}} \right)}  + \sum\limits_{j = 1}^{N_s} {{\mathbf{W}_{s,j}}}  + {\mathbf{\Sigma} _s}} \right]}\\ \notag
&\text{s.t.}\ \\
& C1:{R_{P,m,i}} \ge {\gamma _{P,m,i}},i \in {{\cal{N}}_{P,m}},m \in \cal{M},\\
&  C2:{R_{S,j}} \ge {\gamma _{S,j}},j \in {{\cal{N}}_s},\\
& C3:\text{Tr}\left\{\left( {\sum\limits_{j = 1}^{N_s} {{\mathbf{W}_{s,j}}}  + {\mathbf{\Sigma} _s}} \right){\mathbf{F}_{S,m,i}}\right\} \le {\Upsilon _{m,i}},i \in {{\cal{N}}_{P,m}},\\
&  C4:{\Phi _{E,{A_1}}} \ge {\zeta _{E,{A_1}}},k \in {{{\cal{K}}{_{m}}}},m \in \cal{M},\\
&  C5:{\Phi _{E,{A_2}}} \ge {\zeta _{E,{A_2}}},l \in {\cal{K}}_s,\\
&  C6:\text{Rank}\left(\mathbf{W}_{p,m}\right)=1, \text{Rank}\left(\mathbf{W}_{s,j}\right)=1
\end{align}
\end{subequations}
where ${\gamma _{P,m,i}}$ and $\gamma _{S,j}$ are the minimum secrecy rate requirements of the $i$th PU in the $m$th cluster and of the $j$th SU; $\Upsilon _{m,i}$ is the maximum tolerable interference power of the $i$th PU in the $m$th cluster; $\zeta _{E,{A_1}}$ and $\zeta _{E,{A_2}}$ are the minimum EH requirements of EHRs in the primary and the secondary network. Due to constraints $C1$, $C2$ and $C6$, $\text{P}_{{1}}$ is non-convex and difficult to be solved. In order to solve this problem, a suboptimal scheme based on semidefinite relaxation (SDR) and
successive convex approximation (SCA) is proposed.
\subsection{Suboptimal Solution}
To address constraint $C1$, auxiliary variables $\tau_m$, $m \in \cal{M}$, are introduced. $C1$ can be equivalently
expressed as
\begin{subequations}
\begin{align}\label{27}\
&\log \left\{ {\frac{{{\Gamma _{P,m,i}}}}{{\left[ {{\Gamma _{P,m,i}} - \text{Tr}\left( {{\mathbf{W}_{p,m}}{\mathbf{H}_{P,m,i}}} \right)} \right]{\tau _m}}}} \right\} \ge {\gamma _{P,m,i}},\\
&\log \left\{ {\frac{{{\Gamma _{E,m,k}} + \sigma _{E,m,k}^2}}{{\left[ {{\Gamma _{E,m,k}} - \text{Tr}\left( {{\mathbf{W}_{p,m}}{\mathbf{G}_{E,m,k}}} \right) + \sigma _{E,m,k}^2} \right]{\tau _m}}}} \right\} \le 1,
\end{align}
\end{subequations}
where $k \in {{{\cal{K}}{_{m}}}}$ and $m \in \cal{M}$. Using SCA, constraints given by $\left(6\rm{a}\right)$ and $\left(6\rm{b}\right)$ can be approximated as $\left(7\right)$ and $\left(8\right)$
\begin{subequations}
\begin{align}\label{27}\
&\exp \left( {{\alpha _{P,m,i}} + {\beta _m} - {\lambda _{P,m,i}}} \right) \le {2^{ - {\gamma _{P,m,i}}}},\\ \notag
&{\Gamma _{P,m,i}} - \text{Tr}\left( {{\mathbf{W}_{p,m}}{\mathbf{H}_{P,m,i}}} \right)  \\
&\ \ \ \ \ \ \ \ \ \ \ \ \ \ \ \ \ \ \le\exp \left( {{{\widetilde \alpha }_{P,m,i}}} \right)\left( {{\alpha _{P,m,i}} - {{\widetilde \alpha }_{P,m,i}} + 1} \right),\\
&{\tau _m} \le \exp \left( {{{\widetilde \beta }_m}} \right)\left( {{\beta _m} - {{\widetilde \beta }_m} + 1} \right),\\
&{\Gamma _{P,m,i}} \ge \exp \left( {{\lambda _{P,m,i}}} \right),
\end{align}
\end{subequations}
\begin{subequations}
\begin{align}\label{27}\
&\exp \left( {{\mu _{E,m,k}} - {\rho _{E,m,k}} - {\delta _m}} \right) \le 1,\\
&{\Gamma _{E,m,k}} + \sigma _{E,m,k}^2 \le \exp \left( {{{\widetilde \mu }_{E,m,k}}} \right)\left( {{\mu _{E,m,k}} - {{\widetilde \mu }_{E,m,k}} + 1} \right),\\
&{\Gamma _{E,m,k}} - \text{Tr}\left( {{\mathbf{W}_{p,m}}{\mathbf{G}_{E,m,k}}} \right) + \sigma _{E,m,k}^2 \ge \exp \left( {{\rho _{E,m,k}}} \right),\\
&{\tau _m} \ge \exp \left( {{\delta _m}} \right),
\end{align}
\end{subequations}
where $\alpha _{P,m,i}$, ${\beta _m}$, $\lambda _{P,m,i}$, $\mu _{E,m,k}$, $\rho _{E,m,k}$  and $\delta _m$ are auxiliary variables. ${\widetilde \alpha }_{P,m,i}$, ${\widetilde \beta }_m$ and ${\widetilde \mu }_{E,m,k}$ are approximate values, and they are equal to $\alpha _{P,m,i}$, ${\beta _m}$ and $\mu _{E,m,k}$, respectively when the constraints are tight. Similarly, constraint $C2$ can be approximated as $\left(9\right)$ and $\left(10\right)$. When $j=N_s$, the secrecy rate constraint of the $N_s$th SU can be given as
\begin{subequations}
\begin{align}\label{27}\
&\exp \left( {{\alpha _{s,{N_s}}} + {\beta _{s,{N_s}}} - {\lambda _{s,{N_s}}}} \right) \le {2^{ - {\gamma _{S,{N_s}}}}},\\ \notag
&{\Gamma _{S,{N_s}}} - \text{Tr}\left( {{\mathbf{W}_{s,{N_s}}}{\mathbf{H}_{S,{N_s}}}} \right) \\
&\ \ \ \ \ \ \ \ \ \ \ \ \ \ \ \  \le \exp \left( {{{\widetilde \alpha }_{s,{N_s}}}} \right)\left( {{\alpha _{s,{N_s}}} - {{\widetilde \alpha }_{s,{N_s}}} + 1} \right),\\
&{\tau _{S,{N_s}}} \le \exp \left( {{{\widetilde \beta }_{s,{N_s}}}} \right)\left( {{\beta _{s,{N_s}}} - {{\widetilde \beta }_{s,{N_s}}} + 1} \right),\\
&{\Gamma _{S,{N_s}}} \ge \exp \left( {{\lambda _{s,{N_s}}}} \right),\\
&\exp \left( {{\mu _{E,l}} - {\rho _{s,l}} - {\omega _{S,{N_s}}}} \right) \le 1,l \in {\cal{K}}_s,\\
&{\Lambda _{E,l,{N_s}}} \le \exp \left( {{{\widetilde \mu }_{E,l}}} \right)\left( {{\mu _{E,l}} - {{\widetilde \mu }_{E,l}} + 1} \right),\\
&{\Lambda _{E,l,{N_s}}} - \text{Tr}\left( {{\mathbf{W}_{s,{N_s}}}{\mathbf{G}_{E,l}}} \right) \ge \exp \left( {{\rho _{s,l}}} \right),\\
&{\tau _{S,{N_s}}} \ge \exp \left( {{\omega _{S,{N_s}}}} \right),
\end{align}
\end{subequations}
where $\alpha _{s,{N_s}}$, $\beta _{s,{N_s}}$, $\lambda _{s,{N_s}}$, $\mu _{E,l}$, $\rho _{s,l}$  and $\omega _{S,{N_s}}$ are auxiliary variables; ${\widetilde \alpha }_{s,{N_s}}$, ${\widetilde \beta }_{s,{N_s}}$ and ${\widetilde \mu }_{E,l}$ are approximate values, and they are equal to $\alpha _{s,{N_s}}$, $\beta _{s,{N_s}}$ and $\mu _{E,l}$, respectively when the constraints are tight. When $j=1,2,\cdots,N_s-1$, the secrecy rate constraint of the $j$th SU can be given as
\begin{subequations}
\begin{align}\label{27}\
&{\kappa _j} - {\omega _j}{2^{{\gamma _{S,j}}}} \ge 0,\\
&\exp \left( {{\alpha _{s,j,z}} + {\xi _{s,j}} - {\lambda _{s,j,z}}} \right) \le 1,z \in \left\{ {j,j + 1,{N_s}} \right\},\\
&{\Lambda _{S,j,z}} - \text{Tr}\left( {{\mathbf{W}_{s,j}}{\mathbf{H}_{S,z}}} \right) \le \exp \left( {{{{\widetilde \alpha }_{s,j,z}}}} \right)\left( {{\alpha _{s,j,z}} - {{\widetilde \alpha }_{s,j,z}} + 1}\right),\\
&{\kappa _j} \le \exp \left( {{{\widetilde \xi }_{s,j}}} \right)\left( {{\xi _{s,j}} - {{\widetilde \xi }_{s,j}} + 1} \right),\\
&{\Lambda _{S,j,z}} \ge \exp \left( {{\lambda _{s,j,z}}} \right),\\
&\exp \left( {{\mu _{E,l,j}} - {\rho _{s,l,j}} - {\tau _{S,j}}} \right) \le 1,\\
&{\Lambda _{s,l,j}} \le \exp \left( {{{\widetilde \mu }_{E,l,j}}} \right)\left( {{\mu _{E,l,j}} - {{\widetilde \mu }_{E,l,j}} + 1} \right),\\
&{\Lambda _{s,l,j}} - \text{Tr}\left( {{\mathbf{W}_{s,j}}{\mathbf{G}_{E,l}}} \right) \ge \exp \left( {{\rho _{s,l,j}}} \right),\\
&{\omega _j} \ge \exp \left( {{\tau _{S,j}}} \right),
\end{align}
\end{subequations}
where $\kappa _j$, $\omega _j$, $\alpha _{s,j,z}$, $\xi _{s,j}$, $\lambda _{s,j,z}$, $\mu _{E,l,j}$, $\rho _{s,l,j}$  and $\tau _{S,j}$ denote auxiliary variables; ${\widetilde \alpha }_{s,j,z}$, ${\widetilde \xi }_{s,j}$ and ${\widetilde \mu }_{E,l,j}$ are approximate values and equal to $\alpha _{s,j,z}$, $\xi _{s,j}$ and $\mu _{E,l,j}$, respectively when the constraints are tight. Constraints $C4$ and $C5$ can be equivalently expressed as
\begin{align}\label{27}\ \notag
&{\Gamma _{E,A}}\\
&\ge {b_{E,A}} - \frac{1}{{{a_{E,A}}}}\ln \left\{ {\frac{{P_{E,A}^{\max }}}{{{\zeta _{E,A}}\left( {1 - {\Psi _{E,A}}} \right) + P_{E,A}^{\max }{\Psi _{E,A}}}} - 1} \right\}.
\end{align}
Based on $\left(7\right)$ and $\left(11\right)$, using SDR, $\text{P}_{{1}}$  can be solved by iteratively solving $\text{P}_{{2}}$, given as
\begin{subequations}
\begin{align}\label{27}\
 &\text{P}_{{2}}: \ \mathop {\min }\limits_{\Xi} \ {\text{Tr}\left[ {\sum\limits_{m = 1}^M {\left( {{\mathbf{W}_{p,m}} + {\mathbf{\Sigma} _{p,m}}} \right)}  + \sum\limits_{j = 1}^{N_s} {{\mathbf{W}_{s,j}}}  + {\mathbf{\Sigma} _s}} \right]}\\
&\text{s.t.}\ \left(7\right)-\left(11\right),
\end{align}
\end{subequations}
where $\Xi$ is the set including all optimization variables and auxiliary variables.  $\text{P}_{{2}}$ is convex and can be efficiently solved by using the software \texttt{CVX} \cite{F. Zhou2}. Algorithm 1 is given to solve $\text{P}_{{1}}$. The details of Algorithm 1 are provided in Table 1. where $P_{opt}^n$ denotes the minimum total transmission power at $n$th iteration.

\begin{table}[htbp]
\begin{center}
\caption{The SCA-based algorithm}
\begin{tabular}{lcl}
\\\toprule
$\textbf{Algorithm 1}$: The SCA-based algorithm for $\text{P}_{{1}}$\\ \midrule
\  1: \textbf{Setting:}\\
\ \  \ $\gamma _{P,m,i}$, $\gamma _{S,j}$ $\Upsilon _{m,i}$, $\zeta _{E,{A_1}}$, $\zeta _{E,{A_1}}$, $i \in {{\cal{N}}_{P,m}}$, $k \in {{{\cal{K}}{_{m}}}}$,  $m \in \cal{M}$\\
\ \  \ $l \in {\cal{K}}_s$ and the tolerance error $\varpi$; \\
\  2: \textbf{Initialization:}\\
\ \  \ The iterative number $n=1$, ${\widetilde \alpha }_{P,m,i}^n$, ${\widetilde \beta }_m^n$, ${\widetilde \mu }_{E,m,k}^n$, ${\widetilde \alpha }_{s,{N_S}}^n$, ${\widetilde \beta }_{s,{N_S}}^n$,\\
\ \  \ ${\widetilde \mu }_{E,l}^n$, ${\widetilde \alpha }_{s,j,z}^n$, ${\widetilde \xi }_{s,j}^n$ and ${\widetilde \mu }_{E,l,j}^n$ and $P_{opt}^n$; \\
\  3: \textbf{Repeat:}\\
 \   \ \ \ solve $\text{P}_{\textbf{2}}$ by using \texttt{CVX} for the given approximate values; \\
\ \ \ \ \ obtain ${\widetilde \alpha }_{P,m,i}^{n+1}$, ${\widetilde \beta }_m^{n+1}$, ${\widetilde \mu }_{E,m,k}^{n+1}$, ${\widetilde \alpha }_{s,{N_S}}^{n+1}$, ${\widetilde \beta }_{s,{N_S}}^{n+1}$, ${\widetilde \mu }_{E,l}^{n+1}$,\\
\ \  \ \ \ ${\widetilde \alpha }_{s,j,z}^{n+1}$, ${\widetilde \xi }_{s,j}^{n+1}$ and ${\widetilde \mu }_{E,l,j}^{n+1}$ and $P_{opt}^{n+1}$; \\
 \ \ \ \ \ if $\text{Rank}\left(\mathbf{W}_{p,m}\right)=1$ and $ \text{Rank}\left(\mathbf{W}_{s,j}\right)=1 $\\
  \ \ \ \ \ \ Obtain optimal $\mathbf{W}_{p,m}$ and $\mathbf{W}_{s,j}$;  \\
 \ \ \ \ \ else \\
 \ \ \ \ \ \ Obtain suboptimal $\mathbf{W}_{p,m}$ and $\mathbf{W}_{s,j}$;   \\
 \ \ \ \ \ end \\
 \ \ \ \ \ update the iterative number $n=n+1$;  \\
 \ \ \ \ \ calculate the total transmit power $P_{opt}^n$;  \\
\ \ \ \ \ if $\left|P_{opt}^n-P_{opt}^{\left(n-1\right)}\right|\leq \varpi$ \\
\ \ \ \ \  \  break;\\
\ \ \ \ \ end;\\
\  4: \textbf{Obtain resource allocation:}\\
 \ \ \ \ \ \ $\mathbf{W}_{p,m}$, $\mathbf{W}_{s,j}$, $\mathbf{\Sigma} _{p,m}$ and $\mathbf{\Sigma} _s$. \\
\bottomrule
\end{tabular}
\end{center}
\end{table}
Algorithm 1 does not guarantee that the optimal beamforming $\mathbf{w}_{p,m}$, $\mathbf{w}_{s,j}$ can be obtained. If $\mathbf{W}_{p,m}$ and $\mathbf{W}_{s,j}$ are of rank-one, the optimal beamforming scheme can be obtained by the eigenvalue decomposition and the obtained eigenvectors are optimal beamforming. If $\mathbf{W}_{p,m}$ and $\mathbf{W}_{s,j}$ are not of rank-one, the suboptimal beamforming vectors can be obtained by using the well-known Gaussian randomization procedure \cite{F. Zhou2}.

\begin{figure*}[!t]
\centering
\subfigure[The minimum transmission power versus the number of EHRs.] {\includegraphics[height=2in,width=2in,angle=0]{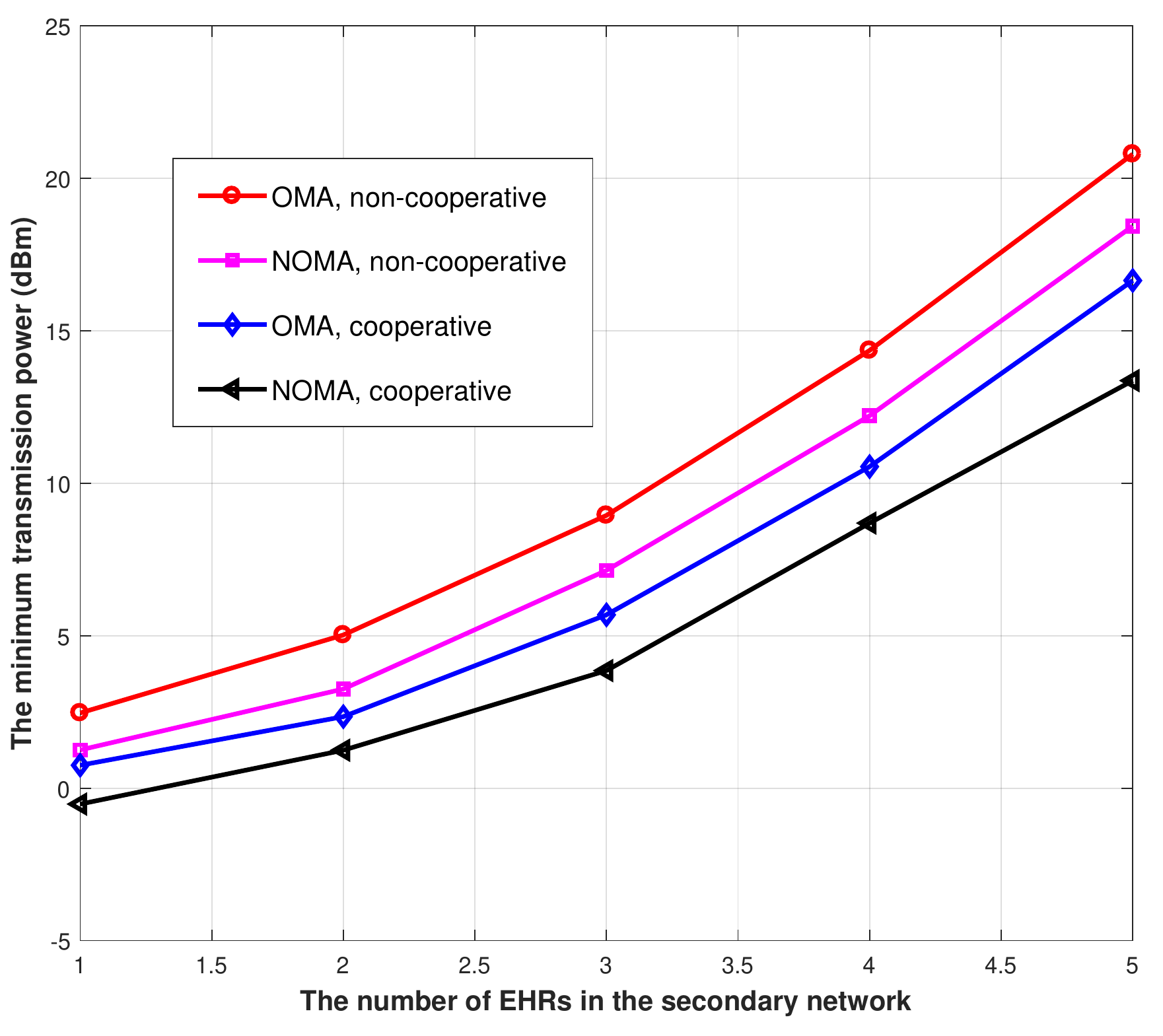}}
\subfigure[The minimum transmission power versus the number of iterations.] {\includegraphics[height=2in,width=2in,angle=0]{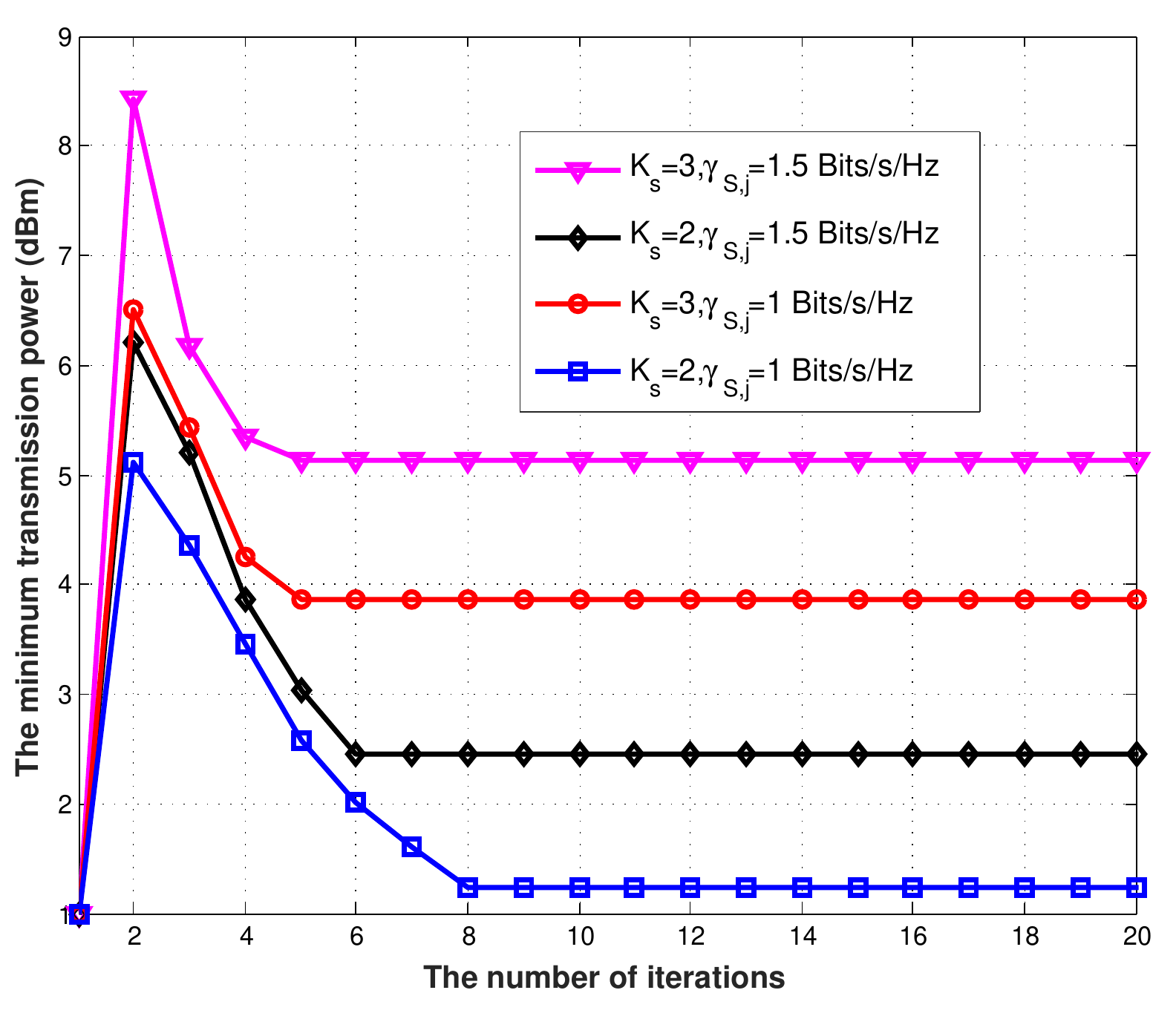}}
\subfigure[The minimum transmission power versus the secrecy rate of PUs.] {\includegraphics[height=2in,width=2in,angle=0]{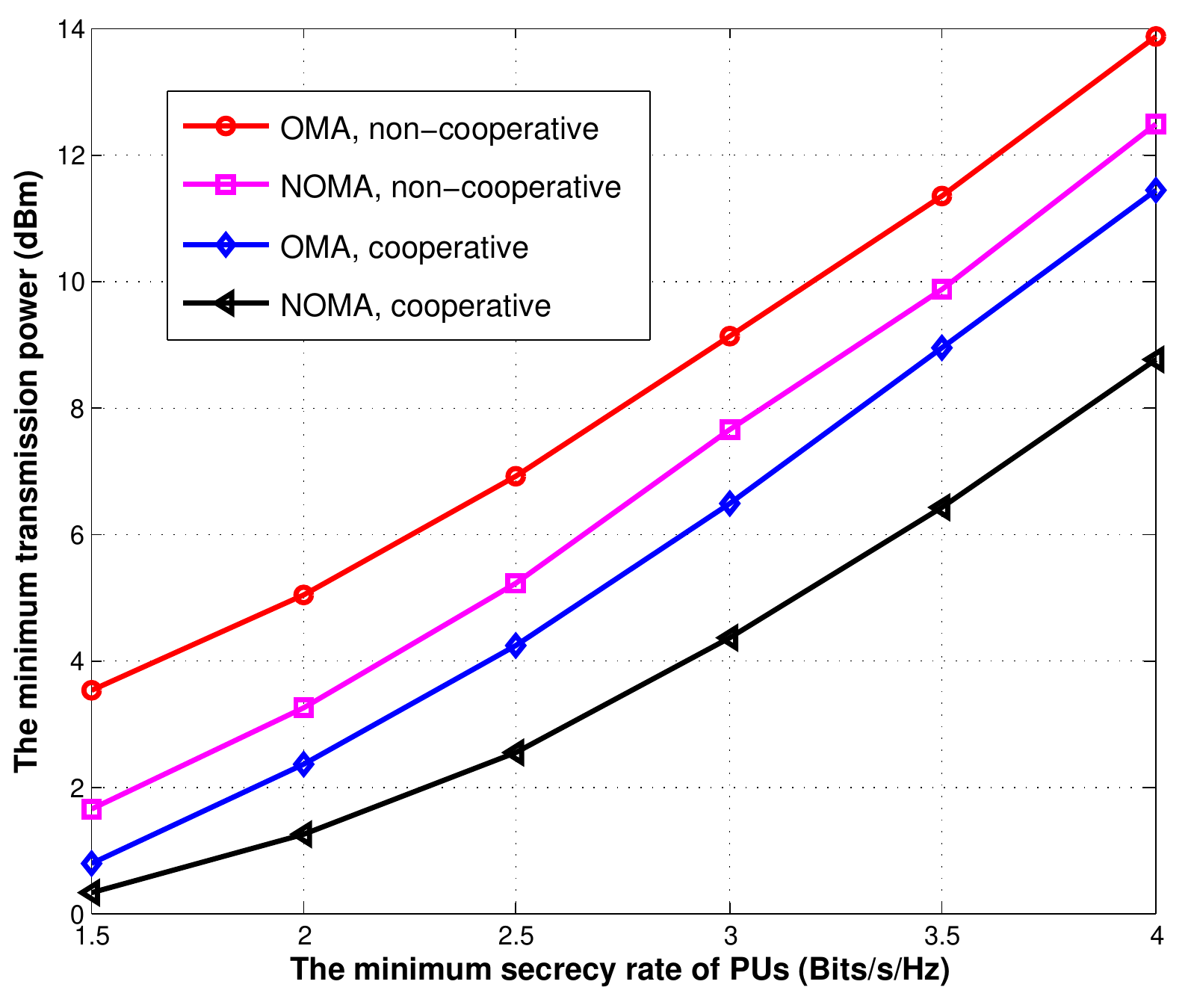}}
\label{fig.1}
\end{figure*}
\section{Simulation Results}
The simulation settings are based on those used  in \cite{E. Boshkovska3} and \cite{E. Boshkovska}. All the involved channels are assumed to be Rayleigh flat fading. The number of channel realizations is $10^4$. The variances of noise at all users and EHRs are $-120$ dBm. The channel distributions are set as: $\mathbf{h}_{P,m,i}\sim{\cal C}{\cal N}\left( {\mathbf{0},2\mathbf{I } }\right)$, $\mathbf{f}_{S,m,i} \sim{\cal C}{\cal N}\left( {\mathbf{0},0.5\mathbf{I } }\right)$, $\mathbf{q}_{P,j} \sim{\cal C}{\cal N}\left( {\mathbf{0},0.5\mathbf{I } }\right)$, $\mathbf{h}_{S,j} \sim{\cal C}{\cal N}\left( {\mathbf{0},2\mathbf{I } }\right)$, $\mathbf{g}_{E,m,k} \sim{\cal C}{\cal N}\left( {\mathbf{0},1.5\mathbf{I } }\right)$, $\mathbf{f}_{E,m,k} \sim{\cal C}{\cal N}\left( {\mathbf{0},0.5\mathbf{I } }\right)$, $\mathbf{q}_{E,l}\sim{\cal C}{\cal N}\left( {\mathbf{0},0.5\mathbf{I } }\right)$ and $\mathbf{g}_{E,l} \sim{\cal C}{\cal N}\left( {\mathbf{0},1.5\mathbf{I } }\right)$. The detailed simulation settings are given in Table III.

\begin{table}[htbp]
 \caption{\label{tab:test}Simulation Parameters}
 \begin{tabular}{l|c|c}
  \midrule
  \midrule
  Parameters & Notation & Typical Values  \\
  \midrule
  \midrule
 Numbers of antennas of the PBS & $N_{P,t}$ & $10$ \\
 Numbers of antennas of the CBS & $N_{S,t}$ & $5$ \\
 Numbers of the clusters & $M$ & $2$ \\
 Numbers of SUs & $N_s$ & $3$ \\
 The maximum harvested power & $P_{E,A}^{\max }$ & $24$ mW \\
 Circuit parameter & $a_{E,A}$ & $1500$ \\
 Circuit parameter & $b_{E,A}$ & $0.0022$ \\
  The minimum secrecy rate of PUs & $\gamma _{P,m,i}$ & $2$ bits/s/Hz \\
 The minimum secrecy rate of SUs& $\gamma _{S,j}$ & $1$ bits/s/Hz \\
 The maximum interference power & $\Upsilon _{m,i}$ & $10$ mW \\
 The minimum EH of EHRs in set $A_1$& $\zeta _{E,A_1}$ & $15$ mW\\
 The minimum EH of EHRs in set $A_2$& $\zeta _{E,A_2}$ & $5$ mW\\
 The tolerance error & $\varpi$ & $10^{-4}$ \\
\midrule
\midrule
 \end{tabular}
\end{table}
Fig. 2(a) shows the minimum transmission power versus the number of EHRs in the secondary network. It can be  seen that the minimum transmission power consumed without the cooperative jamming scheme is larger than that consumed with our proposed cooperative jamming scheme. The reason is that our proposed cooperative jamming scheme is efficient for secure communication. As shown in Fig. 2(b), it only needs several iterations to converge to the minimum transmission power. This indicates the efficiency of our proposed algorithm. Fig. 2(c) is given to further verify that our proposed cooperative scheme is beneficial to improve the security of NOMA CRNs using SWIPT. It is also seen from Fig. 2(a) and Fig. 2(b) that NOMA outperforms OMA (time division multiple access is used) in terms of the power consumption.

\section{Conclusion}
Secure communication was studied in a MISO NOMA CRN using SWIPT where a practical non-linear EH model was considered. An AN-aided cooperative jamming scheme was proposed to improve the security of both the primary and secondary network. The total transmission power was minimized by jointly optimizing the transmission beamforming and the AN covariance matrix. It was shown that our proposed cooperative jamming scheme is efficient to achieve secure communication. Simulation results also show that  the performance achieved by using NOMA is better than that obtained by using OMA in terms of the power consumption.


\begin{thebibliography}{20}
\bibitem{J. G. Andrews}
F. Zhou, \emph{et al.}, \lq\lq State of the art, taxonomy, and open issues on NOMA in cognitive radio networks?,\rq\rq \ \emph{IEEE Wireless Commun.}, to appear, 2017.
\bibitem{S. Haykin}
R. Q. Hu and Y. Qian, \lq\lq An energy efficient and spectrum efficient wireless heterogeneous network framework for 5G systems,\rq\rq \ \emph{IEEE Commun. Mag.}, vol.52, no.5, pp.94-101, May 2014.
\bibitem{Z. Ding}
F. Zhou, \emph{et al.}, \lq\lq Energy-efficient NOMA enabled  heterogeneous cloud radio access networks,\rq\rq \ \emph{IEEE Network}, to be published, 2017.
\bibitem{R. Q. Hu1}
L. Wei, R. Q. Hu, \emph{et al.}, \lq\lq Enabling device-to-device communications underlaying cellular networks: challenges and research aspects,\rq\rq \ \emph{IEEE Commun. Mag.}, vol.52, no.6, pp.90-96, June 2014.
\bibitem{Y. Liu}
Y. Liu, \emph{et al.}, \lq\lq Nonorthogonal multiple access in large-scale underlay cognitive radio networks,\rq\rq \ \emph{IEEE Trans. Veh. Technol.}, vol. 65, no. 12, pp. 10152-10157, Dec. 2016.
\bibitem{Z. Zhang}
Z. Zhang, \emph{et al.}, \lq\lq Downlink and uplink non-orthogonal multiple access in a dense wireless network,\rq\rq \ \emph{IEEE J. Sel. Areas Commun.}, to be published, 2017.
\bibitem{X. Lu}
X. Lu, \emph{et al.}, \lq\lq Wireless networks with RF energy harvesting: A contemporary survey,\rq\rq \ \emph{IEEE Commun. Surveys Tuts.}, vol. 17, pp. 757-789, Second Quarter, 2015.
\bibitem{F. Zhou2}
F. Zhou, \emph{et al.}, \lq\lq Robust AN-Aided beamforming and power splitting design for secure MISO cognitive radio with SWIPT,\rq\rq \ \emph{IEEE Trans. Wireless Commun.}, vol. 16, no. 4, pp. 2450-2464, April 2017.
\bibitem{E. Boshkovska3}
E. Boshkovska, \emph{et al.}, \lq\lq Secure SWIPT networks based on a non-linear energy harvesting model,\rq\rq \ in \emph{Proc. IEEE WCNC 2017},  San Francisco, CA, USA, 2017.
\bibitem{Y. Zhang1}
Y. Zhang, \emph{et al.}, \lq\lq Secrecy sum rate maximization in non-orthogonal multiple access,\rq\rq \ \emph{IEEE Commun. Lett.}, vol. 20, no. 5, pp. 930-933, 2016.
\bibitem{Y. Zou1}
L. Wei, R. Q. Hu, Y. Qian, G. Wu, \lq\lq Enabling device-to-device communications underlaying cellular networks: challenges and research aspects,\rq\rq \ \emph{IEEE Commun. Mag.}, vol.52, no.6, pp.90-96, June 2014.
 \bibitem{X. Chen}
V. Nguyen, \emph{et al.}, \lq\lq Enhancing PHY security of cooperative cognitive radio communications,\rq\rq \ \emph{IEEE Trans. Cogn. Net.},  to appear, 2017.
\bibitem{Z. Chu}
Z. Chu, \emph{et al.}, \lq\lq Simultaneous wireless information power transfer for MISO secrecy channel,\rq\rq \ \emph{IEEE Trans. Vehicular Technol.}, vol. 65, no. 9, pp. 6913-6925, Sept. 2016.
\bibitem{Y. Li}
Y. Li, \emph{et al.}, \lq\lq Secure beamforming in downlink MISO nonorthogonal multiple access systems,\rq\rq \ \emph{IEEE Trans.
Veh. Technol.}, to appear, 2017.
\bibitem{M. Tian}
M. Tian, \emph{et al.}, \lq\lq Secrecy sum rate optimization for downlink MIMO non-orthogonal multiple access systems,\rq\rq \ \emph{IEEE Signal Process. Lett.}, to be published, 2017.
\bibitem{B. He}
B. He, \emph{et al.}, \lq\lq On the design of secure non-orthogonal multiple access systems,\rq\rq \ \emph{IEEE J. Sel. Areas Commun.}, to be published, 2017.
\bibitem{D. W. K. Ng1}
D. W. K. Ng, \emph{et al.}, \lq\lq Multi-objective resource allocation for secure communication in cognitive radio networks with wireless information and power transfer,\rq\rq \ \emph{IEEE Trans. Veh. Technol.}, vol. 20, no. 2, pp. 328-331, Feb. 2016.
\bibitem{C. Xu}
C. Xu, \emph{et al.}, \lq\lq Robust transceiver design for wireless information and power transmission in underlay MIMO cognitive radio networks,\rq\rq \ \emph{IEEE Commun. Lett.}, vol. 18, no. 9, pp. 1665-1668, Sept. 2014.
\bibitem{E. Boshkovska}
E. Boshkovska, \emph{et al.}, \lq\lq Practical nonLinear energy harvesting model and resource allocation for SWIPT systems,\rq\rq \ \emph{IEEE Commun. Lett.}, vol. 19, pp. 2082-2085, Dec. 2015.
\bibitem{E. Boshkovska2}
E. Boshkovska, \emph{et al.}, \lq\lq Robust resource allocation for MIMO wireless powered communication networks based on a non-linear EH model,\rq\rq \ \emph{IEEE Trans. Commun.}, vol. 65, no. 5, pp. 1984-1999, May 2017.
\bibitem{K. Xiong}
K. Xiong, \emph{et al.}, \lq\lq Rate-energy region of SWIPT for MIMO broadcasting under non-linear energy harvesting model,\rq\rq \ \emph{IEEE Trans. Wireless Commun.}, to be published, 2017.
\end{thebibliography}
\end{document}